\documentclass[article,a4paper,preprintnumbers,11pt]{revtex4-2}

\usepackage{amsmath,amssymb,amsfonts,bbm,amsthm, changes, graphicx, dsfont, verbatim, appendix}
\usepackage[T1]{fontenc}  
\DeclareMathAlphabet{\mathcal}{OMS}{cmsy}{m}{n}

\usepackage[hyperindex, breaklinks]{hyperref}
\hypersetup{
     colorlinks=true,        % false: boxed links; true: colored links
     citecolor=blue,    % color of links to bibliography      
     filecolor=blue,      		% color of file links
     urlcolor=blue,           	% color of external links    
    runcolor=cyan,
}

\usepackage[left=2.5cm,right=2.5cm,top=2.5cm,bottom=2.5cm]{geometry}

\usepackage{orcidlink}
\usetikzlibrary{decorations.pathreplacing}

\theoremstyle{definition}
\newtheorem{Lemma}{Lemma}
\newtheorem{Definition}[Lemma]{Definition}
\newtheorem{Theorem}[Lemma]{Theorem}

\begin{document}

\newcommand{\B}{\textrm{Bott}}
\newcommand{\T}{\textrm{Tr}}
\newcommand{\PP}{P^\bot}
\newcommand{\bfk}{\mathbf{k}}
\newcommand{\bfx}{\mathbf{x}}
\newcommand{\e}{e}
\newcommand{\I}{\mathds{1}}
\newcommand{\id}{\mathds{1}}
\newcommand{\Or}{\matcal{O}}
\newcommand{\Tr}{\mathrm{Tr}}
\newcommand{\lam}{\lambda}

%% Text
\newcommand{\eq}{Eq.}%No extra space when used with reftex (->auto ~)
\newcommand{\eqs}{Eqs.}%No extra space when used with reftex (->auto ~)
\newcommand{\cf}{\textit{cf. }}%adv : that is to say; in other words
\newcommand{\ie}{\textit{i.e. }}%adv : that is to say; in other words
\newcommand{\eg}{\textit{e.g. }}%[syn: f.eks., for example, for instance]
\newcommand{\etal}{\emph{et al.}}
\newcommand{\tx}{\phi_x}
\newcommand{\ty}{\phi_y}

\def\i{\mathrm{i}}

%\crefname{section}{Section}{Sections}
% 
% \renewcommand{\thesection}{\arabic{section}}
% \renewcommand{\thesubsection}{\thesection.\arabic{subsection}}
% \renewcommand{\thesubsubsection}{\thesubsection.\arabic{subsubsection}}

\title{On the Bott index of unitary matrices on a finite torus}

\author{Daniele Toniolo\,\orcidlink{0000-0003-2517-0770}}

\affiliation{Department of Computer Science, University College London, United Kingdom}
\affiliation{Department of Physics and Astronomy, University College London, United Kingdom}

\email{danielet@alumni.ntnu.no}
\email{d.toniolo@ucl.ac.uk}

\date{\today}

%----------------------------------------------------------------------------------------------
%                                ABSTRACT
%----------------------------------------------------------------------------------------------

\begin{abstract}
This article reviews the foundations of the theory of the Bott index of a pair of unitary matrices in the context of condensed matter theory, as developed by Hastings and Loring \cite{Hastings_Loring:2010,Hastings_Loring:2011}, providing a novel proof of the equality  with the Chern number. The Bott index is defined for a pair of unitary matrices, then extended to a pair of invertible matrices and homotopic invariance of the index is proven. An insulator defined on a lattice on a 2-torus, that is a rectangular lattice with periodic boundary conditions, is considered and a pair of quasi-unitary matrices associated to this physical system are introduced. It is shown that their Bott index is well defined and the connection with the transverse conductance, the Chern number, is established proving the equality of the two quantities.
\end{abstract}

\maketitle

%---------------------------------------------------------------------------
%                               INTRODUCTION
%---------------------------------------------------------------------------

\section{Introduction}

The integer quantization of the transverse (Hall) conductance (IQHE) of a two dimensional 
electron gas 
under an external perpendicular magnetic field has been experimentally discovered in 1980 
\cite{von_Klitzing:1980}, the fractional quantization (FQHE) a couple of years later 
\cite{Stormer:1982}. The 
theoretical analysis of these phenomena has never stopped since. Initial landmarks have been 
established by Laughlin \cite{Laughlin:1981}, Halperin \cite{Halperin:1982} and Thouless 
{\it et al.} \cite{Thouless:1982}. Different schools of thought originated to explain these phenomena: who
focused on the two-dimensional bulk aspects of the sample \cite{Thouless_review:1994}; who 
stressed the relevance of the one-dimensional 
edge \cite{Wen_review:1992}, who analyzed the interplay
between bulk and edge-physics \cite{Frohlich_Studer_review:1993}. The attention to a realistic 
geometrical setting is particularly relevant in the approach of Buttiker \cite{Buttiker:1988}, 
while a rigorous treatment of the strong disorder needed for the quantization of the conductance is 
central in the work of Bellissard, summarized in \cite{Bellissard:1994}. The 
initial sections of \cite{Bellissard:1994} can be used as an introduction to the 
IQHE. Another line of research on the mathematical-physics aspects of 
IQHE is due to Avron, Seiler and Simon \cite{Avron_Seiler:1985,Avron_Seiler_Simon:1994}. 
Haldane in 1988 formulated a lattice model with localized magnetic flux over the corners of a 
honeycomb lattice but with total magnetic flux per plaquette equal to zero. This model manifests a quantized transverse 
conductance \cite{Haldane:1988} and nowadays is called Chern insulator. The Haldane model has been relevant for 
the theoretical formulation \cite{Kane_1_2005,Kane_2_2005} and experimental discovery 
\cite{Konig:2007,Knez_Rice_2014} of the topological insulators. 
Two relatively recent rigorous works, among others, on the nature of the invariants describing topological insulators in two and higher dimensions are \cite{Fiorenza_2016,Schulz_Baldes_2016}. A rigorous discussion of the topology of one-dimensional systems with open boundary conditions has been very recently provided by the authors of \cite{Tauber_2022}. 

The quantization of the Hall conductance on a torus geometry, meaning that periodic boundary 
conditions are imposed on a two dimensional rectangular sample, is determined by a topological invariant 
called Chern number. This has been showed for the first time in the work of Thouless {\it et al.} \cite{Thouless:1982}, the 
emergence of the Chern number has then been made explicit by Kohmoto in \cite{Kohmoto:1985}. Also early 
contributions have been made in \cite{Avron:1983} and \cite{Simon:1983}. 

Hastings and Loring in a set of articles \cite{Loring_Hastings:2010,Hastings_Loring:2010,Hastings_Loring:2011} used  mathematical tools including non commutative topology, 
C*-algebras and K-theory to rigorously search for the topological invariants of the ten 
Altland and Zirnbauer symmetry classes \cite{Altland_Zirnbauer:1997} in a way that is also  
relevant for numerical computations. The program of classification of topological invariants of 
Fermi systems according to their symmetries and dimensionality started with the works of Qi {\it et al.} 
\cite{Qi_Zhang:2008}, Kitaev \cite{Kitaev:2009} and Ryu {\it et al.} \cite{Ryu:2010}.

One of the motivations for this article is to review the foundations of the theory of the Bott index of a pair of unitary matrices in the context of condensed matter theory, as developed by Hastings and Loring \cite{Hastings_Loring:2010,Hastings_Loring:2011}, and in particular showing the equivalence with the Chern number, with a novel proof of the equality of the two indices, providing throughout new proofs that make this work self-contained. 

In the physics literature the Bott index, following \cite{Hastings_Loring:2010,Hastings_Loring:2011}, has been employed to characterize several topological phenomena, that goes from time-reversal invariant systems \cite{Huang_2018_2}, time-dependent systems \cite{Refael:2015, Ge_Rigol:2017, Toniolo:2018}, quasi periodic systems \cite{Huang_2018,Loring_2019_2, Nielsen_2020,Yoshii_2021} and ferromagnetic systems \cite{Brataas_2020}. The former list is  not exhaustive.

The structure of this article is as follows: in section \ref{section_physical_setting} the physical setting is presented: a lattice on a two-torus, that is a finite rectangular lattice with periodic boundary conditions, is considered and an insulator is defined on it. This is modeled by the Fermi projection that fills the eigenstates of a short-ranged, bounded, gapped, single-particle Hamiltonian below a spectral gap. The Hilbert space where the Hamiltonian is defined is finite dimensional. The most important results of this section are the bounds \eqref{H_bound} and \eqref{projectorcomm}. The section \ref{section_Bott_index} defines and discusses the Bott index of a pair of unitary matrices according to \cite{Exel_Loring_1989,Exel_Loring_1991,Loring:2014,Loring_2015}, with the important generalization in subsection \ref{sub_Bott_inv} to a pair of invertible matrices and the proof of homotopy invariance in subsection \ref{sub_hom_inv}.
Section \ref{vanishing_index} gives a sufficient condition for the vanishing of the Bott index. 
Section \ref{section_equivalence} inspired by the approach of \cite{Hastings_Loring:2011} in \ref{Bott-Chern-approx} proves that the Bott index approximates the Chern number. A novel proof of the equality of the two indices is given in subsection \ref{Bott-Chern-exact} through a mapping to a differential equation following an analogous proof for the infinite two-dimensional case recently presented in \cite{Toniolo_Bott_2021}. Three perspectives for future developments are given in section \ref{section_discussion}.

I state here the main new result of this work that is an equality among the Bott index of a pair of unitary matrices related to a Hamiltonian describing an insulator, on a 2-torus, as specified in definition \ref{Ham_single}, and the Chern number of its Fermi projection. The proof of the theorem is in section \ref{Bott-Chern-exact}. Previously this relation has been established only as an approximate equality making use of a pair of quasi-unitary matrices, this approach is described in section \ref{Bott-Chern-approx}.

{\bf Theorem} \ref{theo2}. Given a Hamiltonian $ H $ as in definition \ref{Ham_single} and its Fermi projection $ P $, the unitary matrices
\begin{equation}
 e^{2 \pi i P \frac{X}{L} P } \hspace{2mm} , \hspace{2mm} e^{2 \pi i P \frac{Y}{L} P }  \nonumber
\end{equation}
 have a well defined Bott index that satisfies: 
\begin{equation} 
 \textrm{Bott}\left( e^{2 \pi i P \frac{X}{L} P },e^{2 \pi i P \frac{Y}{L} P } \right) = 2\pi i \mathrm{Tr} \left[ P \frac{X}{L}P,P \frac{Y}{L}P \right] = -\frac{4\pi}{L^2}\mathrm{Im}\mathrm{Tr} \left(P[X,P][Y ,P] \right)= \mathrm{Ch}(P) \nonumber
\end{equation}

\section{Physical setting} \label{section_physical_setting}

The physical system under investigation is an insulator comprised of free fermions on a lattice on a two-torus (that is a rectangular lattice with periodic boundary conditions) filling up the energy levels of a single particle Hamiltonian that is short-ranged, bounded and gapped. The system's fermions have in general $N$ internal degrees of freedom. The system admits weak disorder meaning that the Hamiltonian maintains a spectral gap, the disorder is also supposed to be compatible with the periodic boundary conditions. The presence of disorder makes the concept of Brillouin zone ill defined, therefore I do not refer to it. The Hamiltonian has no extra symmetries. The first 
application of the Bott index in condensed matter theory has been provided in \cite{Loring_Hastings:2010}. 

This section shows that from the properties of the Hamiltonian, short-range, bounded and gapped, 
two important estimates on norms of commutators follow, that in turn allow the introduction of two suitable quasi-unitary and unitary matrices whose  Bott index will be evaluated in section \ref{section_equivalence}.

\begin{Definition} \label{Ham_single}
The single particle Hamiltonian $ H : l^2(\Lambda) \otimes  \mathds{C}^N \rightarrow l^2(\Lambda) \otimes  \mathds{C}^N $, with $ \Lambda $ denoting a lattice on a two-torus, that is a lattice on a rectangle of sides $ L_x $ and $ L_y $ with periodic boundary conditions, 
\begin{equation} 
H= \sum_{l,k=1}^N \sum_{n,m \in \Lambda} H_{n,m,l,k} |n,l \rangle \langle m,k|
\end{equation}
is as follows:
\begin{itemize}
 \item short-ranged with range $ R $, meaning that: $ H_{n,m,l,k}=0 $ when $\mathrm{dist}(n,m) > R $, with $ R \ll L_x$ and $ R \ll L_y$. 
 \item bounded: $ \|H\|  $ is upper bounded by a finite constant independent from the system's size. 
 \item gapped: there exists an energy gap $ \Delta E $ in the spectrum of $ H $ with lower bound unaltered increasing the size of $ \Lambda $.
 \item $ L_x $, $ L_y $, $ R $, $ \| H \| $ and $ \Delta E $ are such that: $ \frac{R \|H\|}{L_x \Delta E} \ll 1 $ and $ \frac{R \|H\|}{L_y \Delta E} \ll 1 $.
\end{itemize}
\end{Definition}

The distance on the lattice is compatible with periodic boundary conditions. To exemplify let us consider a square lattice, with lattice distance $ 1 $ over the rectangle of sides $ L_x $ and $ L_y $ with periodic boundary conditions, then given $ n =(n_x,n_y) $ and $ m=(m_x,m_y) $ with $ \{n_x,m_x\} \in \{0,...,L_x-1\} $ and $ \{n_y,m_y\} \in \{0,...,L_y-1\} $, it is $ \mathrm{dist}(n,m)=\min_{k\in \mathds{Z}}|n_x-m_x+kL_x| + \min_{l\in \mathds{Z}}|n_y-m_y+lL_y| $.

For the sake of simplicity from now on it is assumed $ N=1 $. 

We ``build up'' an insulator out of the single particle Hamiltonian \eqref{Ham_single} filling from the bottom the single particle energy levels till an energy gap of size $ \Delta E $ is reached. 
\begin{Definition}
Given the Hamiltonian $ H $ as defined in \ref{Ham_single} and the chemical potential $ \mu $ fixed within the spectral gap $ \Delta E $, the orthogonal projection $ P := \chi(H\le\mu) $ is called the Fermi projection. 
\end{Definition}
We now introduce well defined position operators on the torus. 
To construct the torus we glue together the opposite sides of a rectangle of linear sizes $ L_x $ 
and $ L_y $. We assign an ordering to the points of the rectangular lattice, such that the  $ i $-th point has coordinates $ (x_i,y_i) $.  We then construct the diagonal matrix $ X $, with elements $ X_{i,j}=x_i 
\delta_{i,j} $, and the corresponding matrix $ Y $, $ Y_{i,j}=y_i \delta_{i,j}$. The matrices $ X $ and $ Y $  have 
$ L_x L_y $ diagonal elements. Note that points that are physically close on the lattice may have corresponding entries in the matrix $ X $ distant from each other, but at most $ L_x $ elements far away. We then define the diagonal unitary matrices that are well defined with respect to periodic boundary conditions, namely $ X \rightarrow X + L_x \mathds{1} $ and $ Y \rightarrow Y + L_y \mathds{1}$:
\begin{equation}
 \exp \left(i\frac{2\pi}{L_x}X\right), \, \exp \left(i\frac{2\pi}{L_y}Y\right) 
\end{equation}
From now on it is set $ L:= L_x = L_y $.

\begin{Lemma} \label{lemma_3}
\begin{equation} \label{H_bound}
 \| [e^{2 \pi i \frac{X}{L}},H] \| \le  \mathcal{O} \left( \frac{R}{L} \|H\| \right)
\end{equation}
and 
\begin{equation} \label{projectorcomm}
\| [e^{2 \pi i \frac{X}{L}},P] \| \le \mathcal{O}\left(\frac{R}{L} \frac{ \|H\| }{\Delta E} \right)
\end{equation}
\end{Lemma}

\begin{proof}
We employ the Holmgren bound for the norm of a bounded operator $ A $, that is:
\begin{equation}
 \| A \| \le \max \left\{ \sup_{m\in \Lambda} \sum_{n \in \Lambda} | \langle m | A |n \rangle | \, , \sup_{n\in \Lambda} \sum_{m \in \Lambda} | \langle m | A |n \rangle | \right\} 
\end{equation}
A proof of this bound can be found, for example, in chapter 16 of \cite{Lax_2002}, for convenience a proof is also presented in appendix \ref{appendix_Holmgren_bound}.
\begin{align}
  \| [e^{2 \pi i \frac{X}{L}},H] \| \le \max \left( \sup_{m\in \Lambda} \sum_{n \in \Lambda} | \langle m | [e^{2 \pi i \frac{X}{L}},H] n \rangle |, m \leftrightarrow n \right) 
\end{align}
We notice that 
\begin{equation}
 \langle m | [e^{2 \pi i \frac{X}{L}},H] | n \rangle = \langle m | (e^{2 \pi i \frac{X}{L}} H - H e^{2 \pi i \frac{X}{L}}) |n \rangle = (e^{2 \pi i \frac{m_x}{L}} - e^{2 \pi i \frac{n_x}{L}})\langle m | H | n \rangle
\end{equation}
Therefore
\begin{align} \label{finite_range}
  \| [e^{2 \pi i \frac{X}{L}},H] \| \le \max \left( \sup_{m\in \Lambda} \sum_{\textrm{dist}(n,m)\le R }  |e^{2 \pi i \frac{m_x}{L}} - e^{2 \pi i \frac{n_x}{L}}| |\langle m | H | n \rangle |, m \leftrightarrow n \right) 
\end{align}
In eq. \eqref{finite_range} we took into account that given a fixed point $ m \in \Lambda $ only the points of the lattice within the range $ R $ contribute to $ \langle m | H | n \rangle $.
We see that 
\begin{equation} \label{mod_L}
 |e^{2 \pi i \frac{m_x}{L}} - e^{2 \pi i \frac{n_x}{L}}| = |e^{2 \pi i \frac{m_x}{L}}\left(1 - e^{2 \pi i \frac{n_x-m_x}{L}}\right) | \le  \frac{2 \pi}{L} \min \{ |n_x-m_x|,L-|n_x-m_x| \} 
\end{equation}
Let us illustrate the bound \eqref{mod_L} with $ m_x=1 $, $ n_x=L-1 $. The points $ (1,y)$ and $ (L-1,y) $ are on the opposite sides of the square lattice but they are close by on the torus because of the periodic boundary conditions and therefore within the range $ R $ of the Hamiltonian.
\begin{equation}
 |e^{2 \pi i \frac{m_x}{L}} - e^{2 \pi i \frac{n_x}{L}}|=|e^{2 \pi i \frac{1}{L}} - e^{2 \pi i \frac{L-1}{L}}|=|e^{2 \pi i \frac{1}{L}} - e^{-2 \pi i \frac{1}{L}}|=|1-e^{-2 \pi i \frac{2}{L}}|\le  \frac{4 \pi}{L}
\end{equation}
It follows that
\begin{equation} 
 \| [e^{2 \pi i \frac{X}{L}},H] \| \le  \mathcal{O} \left( \frac{R}{L} \|H\| \right) \nonumber
\end{equation}
To obtain the bound \eqref{projectorcomm} we start considering, with $ z \in \rho(H) $, and $ A $ any matrix, the equality:
\begin{align}
 & 0=[A,\I]=[A,(H-z\I)(H-z\I)^{-1}]=[A,(H-z\I)](H-z\I)^{-1}+(H-z\I)[A,(H-z\I)^{-1}] \\
 & [A,(H-z\I)^{-1}] = (H-z\I)^{-1}[(H-z\I),A](H-z\I)^{-1} = (H-z\I)^{-1}[H,A](H-z\I)^{-1}
\end{align}
The projection $ P $ on the occupied energy levels, with the loop $ \Gamma $ in the complex plane enclosing them, can be written as
\begin{equation}
P=\frac{1}{2\pi i}\oint_{\Gamma}dz (z\I-H)^{-1}\\
\end{equation}
then:
\begin{align}
 [e^{2\pi i\frac{X}{L}},P] &= \frac{1}{2\pi i}\oint_{\Gamma}dz \left[e^{2\pi i\frac{X}{L}},(z\I-H)^{-1}\right] \\
 &= \frac{1}{2\pi i}\oint_{\Gamma}dz (H-z\I)^{-1}[H,e^{2\pi i\frac{X}{L}}](H-z\I)^{-1} \\
  \| [e^{2\pi i\frac{X}{L}},P] \| &\le \frac{1}{2 \pi}  \| [H,e^{2\pi i\frac{X}{L}}] \| \oint_{\Gamma}  \|(H-z\I)^{-1}\|^2 |dz| \label{partial_bound}
\end{align}
$ \|(H-z\I)^{-1} \| = \mathrm{dist}(z, \sigma(H))^{-1} $

\begin{figure}[h]
\begin{tikzpicture}[scale=0.8]
\useasboundingbox (-7,-3) rectangle (20,6);
\thicklines

\draw[line width=1.5mm,red] (0,1) -- (2.5,1); 
\draw[line width=1.5mm,red] (3.5,1) -- (4.5,1);

\draw[blue] (3,4.5) arc (90:180:3.5cm); 
\draw[blue] (3,-2.5) arc (-90:-180:3.5cm);

\draw[->] (3,-3) -- (3,5);
\draw[->] (-1,1) -- (7,1);

\draw[->,blue] (3,-2.5) -- (3,2.5);
\draw[blue] (3,2.5) -- (3,4.5);

\node[black,above] at (0,4) {$\textrm{the loop}\,\, \Gamma$};
\node[black,above] at (4,4) {$(0,iR)$};
\node[black,above] at (4,-3) {$(0,-iR)$};
\node[black,above] at (1.2,0) {$\sigma(H)$};
\node[black,above] at (4,0) {$\sigma(H)$};

\node[black,above] at (7,0.4) {$\textrm{Re}z$};
\node[black,above] at (2.4,4.7) {$\textrm{i\,Im}z$};

\end{tikzpicture}
\caption{The red stripes enclose the spectrum of $ H $.} 
\label{loop}
\end{figure}

Let us consider the positively oriented loop $ \Gamma $ in the figure \ref{loop}. Along the edge of the loop aligned with the   imaginary-axis of the complex-plane, assuming for simplicity that the energy gap is located around zero, as in figure \ref{loop}, we have that $ \|(H-z)^{-1} \I \|^2 = \mathrm{dist}(z, \sigma(H))^{-2}= 1/\left[(\frac{\Delta E}{2})^2+ (\textrm{Im} z)^2\right] $. Sending $ R \rightarrow  \infty $ the only contribution to the loop-integral comes from the edge along the imaginary-axis, then:
\begin{align}
 \oint_{\Gamma}  \|(H-z\I)^{-1}\|^2 |dz| = \int_{-\infty}^\infty \frac{1}{(\frac{\Delta E}{2})^2+ (\textrm{Im} z)^2} d(\textrm{Im} z) = \frac{2\pi}{\Delta E}
\end{align}
Combining \eqref{H_bound} and \eqref{partial_bound}, this implies:
\begin{equation} 
\| [e^{2 \pi i \frac{X}{L}},P] \| \le \mathcal{O}  \left( \frac{R \|H\|}{L \Delta E } \right)  \nonumber
\end{equation}

\end{proof}

%-----------------------------------------------------------------------------------
%                              BOTT OF A PAIR OF UNITARIES
%-----------------------------------------------------------------------------------

\section{Bott index} \label{section_Bott_index}
The Bott index arose as an index to distinguish pairs of unitary matrices that can be approximated by pairs of commuting unitary matrices from those that cannot. It was established both as a winding number and a K-theoretic invariant in the early works of Exel and Loring \cite{Exel_Loring_1989,Exel_Loring_1991,Exel_1993}. For a discussion of these aspects of the Bott index see \cite{Loring:2014} and references therein.

The Bott index has been employed in the context of condensed matter physics by Hastings 
and Loring in a set of papers \cite{Loring_Hastings:2010,Hastings_Loring:2010,Hastings_Loring:2011}. 
 
In the following the logarithm of a matrix is defined according to the holomorphic (Dunford) functional calculus, for a discussion see for example \cite{Kato_1980}. Denoting $ \rho $ an invertible matrix and $ \Gamma $ a contour enclosing its spectrum but not the origin of the complex plane it is:
\begin{align}
 \log \rho := \frac{1}{2\pi i} \oint_{\Gamma} \log z  (z\I-\rho)^{-1}  dz  
\end{align}
If the spectrum of $ \rho $ does not contain real negative values then the contour $ \Gamma $ is chosen to not intersect the real negative axis of the complex plane and $ \log z $ is the principal logarithm of $ z $. 

\begin{Definition} \label{Bott_unitary_def}
Given two unitary matrices $ U $ and $ V $, such that $ \{-1\}  \notin  \sigma \left( 
UVU^{-1} V^{-1} \right) $, or equivalently such that $\| [U,V ]\| < 2$,  
 their Bott index is defined as:  
\begin{equation} \label{Bott_unitary}
 \mathrm{Bott}(U,V) := \frac{1}{2\pi i}\mathrm{Tr}\log \left(UVU^{-1} V^{-1} \right)
\end{equation}
\end{Definition}

\noindent
{\it Remark}: the equivalence of $ \{-1\}  \notin \sigma \left( UVU^{-1} V^{-1} \right) $ and $\| [U,V ]\| < 2$ follows from:
\begin{equation}
\| UVU^{-1} V^{-1} -\I \|= \|(UV-VU)U^{-1} V^{-1}\|= \| [U,V] \|
\end{equation}

\begin{Lemma}
 The Bott index of two unitary matrices, as in definition \ref{Bott_unitary_def}, is an integer.
\end{Lemma}

\begin{proof}
We denote $ \{e^{i\theta_j}\}, $ with $ \theta_j \in(-\pi, \pi) $, the elements of the spectrum of the unitary matrix $ UVU^{-1} V^{-1} $. From $ \det \left(UVU^{-1} V^{-1} \right) =1 $, it follows that: $ 1 = \prod_j e^{i\theta_j} = e^{i\sum_j \theta_j} $. This implies $ \mathrm{Bott}(U,V) = \frac{1}{2 \pi}\sum_j \theta_j \in \mathds{Z}$.
\end{proof}

It is immediate to see that when $ U $ and $ V $ are commuting their index is vanishing.

%--------------------------------------------------------------------------------------
%                                      BOTT OF INVERTIBLE MATRICES
%--------------------------------------------------------------------------------------

\subsection{The Bott index of two invertible matrices} \label{sub_Bott_inv}
The Bott index of two invertible matrices, $ S $ and $ T $, can be defined similarly as done for unitary matrices. To ensure that $  \log \left( S T S^{-1} T^{-1} \right) $ is well defined, having chosen the branch cut of the logarithm on the real negative axis,  we need that $ \sigma( S T S^{-1} T^{-1} ) $ does not contain any real negative value: $ \sigma( S T S^{-1} T^{-1} ) \bigcap \mathds{R^-} =\emptyset $.  Denoting with $ \lambda_j = |\lambda_j|e^{i\theta_j} $, $ \theta_j \in (-\pi, \pi) $, the set of eigenvalues of $ S T S^{-1} T^{-1} $ we get:
\begin{equation} \label{prod_eign}
 1= \det \left( S T S^{-1} T^{-1} \right) = \prod_j \lambda_j  = \prod_j |\lambda_j| e^{i\theta_j} = \prod_j e^{i\theta_j}  \Rightarrow \frac{1}{2 \pi}\sum_j \theta_j \in \mathds{Z}  
\end{equation}
In equation \eqref{prod_eign} it has been used: $ 1 =  \prod_j \lambda_j = |\prod_j \lambda_j| = \prod_j |\lambda_j| $, implying  $ 0 = \sum_j \log |\lambda_j| $, then
\begin{equation} \label{unitary_part}
 \B( S,T) := \frac{1}{2 \pi i} \T \log \left( S T S^{-1} T^{-1} \right) = \frac{1}{2 \pi i} \sum_j \log \lambda_j =
 \frac{1}{2 \pi i} \sum_j \left( \log |\lambda_j| + i \theta_j \right) = \frac{1}{2 \pi} \sum_j  \theta_j \in  \mathds{Z} 
\end{equation}

\noindent
{\it Remark}:
The unitary matrices are invertible matrices; the reason why two separate definitions are given for their Bott index is that the condition $ \sigma( S T S^{-1} T^{-1} ) \bigcap \mathds{R^-} =\emptyset $ reduces in the unitary case to  $ \{-1\}  \notin \sigma \left( UVU^{-1} V^{-1} \right) $, that is equivalent to $\| [U,V ]\| < 2$, that might be easier to check both analytically or numerically.

%--------------------------------------------------------------------------------------
%                               HOMOTOPIC INVARIANCE
%--------------------------------------------------------------------------------------

\subsection{Homotopy invariance of the Bott index of two invertible matrices} \label{sub_hom_inv}
Homotopy invariance of the Bott index of two unitary matrices has been previously shown by Exel and Loring \cite{Exel_Loring_1989,Exel_Loring_1991,Exel_1993}, in their approach this follows from casting the Bott index as a winding number or a K-theoretic invariant. Here I take a direct approach looking at the derivative of the index.

\begin{Lemma} \label{hom_invariance}

Given two maps $ U(s):[0,1] \rightarrow GL(N,\mathds{C}) $ and $ V(s):[0,1] \rightarrow GL(N,\mathds{C}) $, continuous with respect to the operatorial norm and such that $  \sigma( U(s) V(s) U(s)^{-1} V(s)^{-1} ) \bigcap \mathds{R^-} =\emptyset $.  $ \forall s \in [0,1] $, it holds: 
\begin{equation}
 \B(U(s),V(s))=\B(U(0),V(0))
\end{equation}
\end{Lemma}

\begin{proof}
A continuous path of invertible matrices can be approximated in norm by a differentiable path, \cite{Prodan_Schulz_Baldes} proposition 1.7.2. Let us consider the partial derivatives $ \partial_s U(s) $ and $ \partial_s V(s) $ of such a differentiable path.
\begin{align}
 &\partial_s \T \log (U(s)V(s)U^{-1}(s)V^{-1}(s)) =  \T \left[ \partial_s (U(s)V(s)U^{-1}(s)V^{-1}(s)) (V(s)U(s)V^{-1}(s)U^{-1}(s)) \right] \nonumber \\
 &= \T [ (\partial_s U(s))U^{-1}(s) + U(s)(\partial_s V(s))V^{-1}(s)U^{-1}(s)  
  + U(s)V(s)(\partial_sU^{-1}(s))U(s)V^{-1}(s)U^{-1}(s) \nonumber \\ & \hspace{1.2cm}  + U(s)V(s)U^{-1}(s)(\partial_sV^{-1}(s))V(s)U(s)V^{-1}(s)U^{-1}(s)   ] \\
 &=  \T [(\partial_s U(s)) U^{-1}(s) ] + \T [ (\partial_s V(s))V^{-1}(s)]+ \T [(\partial_s U^{-1}(s))U(s)] + \T[(\partial_sV^{-1}(s))V(s)] \label{hom_inv}
\end{align}
Since $ (\partial_s U^{-1}(s))U(s) = -  U^{-1}(s)\partial_sU(s) $, we obtain: $\partial_s \B(U(s),V(s)) =0 $.
\end{proof}

%-------------------------------------------------------------------------------------
%                                   VANISHING OF THE BOTT INDEX OF A PAIR OF UNITARIES
%-------------------------------------------------------------------------------------

\section{A sufficient condition for the vanishing of the Bott index of a pair of unitaries} \label{vanishing_index}

\begin{Lemma}
Given $ U $ and $ V $ a pair of unitary matrices such that $ \|[U,V]\|<2 $, if $ \|[U,V] \|_1 < 4 $ then $ \textrm{Bott}(U,V)=0 $. $ \| \cdot \|_1 $ denotes the trace norm, namely the sum of the singular values. 
\end{Lemma}

\begin{proof}
The statement follows from the inequality 
\begin{equation} \label{vanish}
 |\textrm{Bott}(U,V)| \le \frac{1}{2\pi} \| \log(UVU^{-1}V^{-1}) \|_1 \le \frac{1}{4} \|[U,V] \|_1
\end{equation}
 Since $ \textrm{Bott}(U,V) $ is an integer, if its modulus  has an upper bound strictly smaller than $ 1 $ then it vanishes. 
 
 Let us prove \eqref{vanish}. Given the set of eigenvalues of $ UVU^{-1}V^{-1} $, $ \{ e^{i\theta_j} \}$ with $ \theta_j \in (-\pi, \pi) $, it holds:
 \begin{align}
 &\frac{1}{2\pi} \| \log(UVU^{-1}V^{-1}) \|_1 = \frac{1}{2\pi} \sum_j |\theta_j|  \label{4} \\
 &\le \frac{1}{4} \sum_j |e^{i\theta_j}-1| = \frac{1}{4} \| UVU^{-1}V^{-1} - \mathds{1} \|_1 = \frac{1}{4} \| (UV - VU) U^{-1}V^{-1} \|_1 \le \frac{1}{4} \| [U,V] \|_1 \label{5}
 \end{align}
The singular values of a normal matrix are the modulus of the eigenvalues; this is used in \eqref{4}. The inequality $ |\theta| \le \frac{\pi}{2} |e^{i\theta}-1| $, with $ \theta \in [-\pi,\pi] $, has been used in \eqref{5}.
\end{proof}

%--------------------------------------------------------------------------------------
%                                              CHERN EQUIVALENCE
%--------------------------------------------------------------------------------------

\section{Equivalence of the Bott index and the Chern number on a finite torus} \label{section_equivalence}

The aim of this section is to introduce a pair of suitable quasi-unitary matrices, given by \eqref{U} below, and a pair of unitary matrices, given by \eqref{U_un}, that arise from the physical system considered in section \ref{section_physical_setting} and prove that their Bott index equals the transverse conductance. This implies that the transverse conductance is an integer, in suitable units.  

In subsection \ref{Bott-Chern-approx} I follow the ideas of Hastings and Loring that realized how to remove the $ \log $ in the expression of the Bott index up to corrections of order $ \frac{R \|H\|}{L \Delta E}  $. In subsection \ref{Bott-Chern-exact} I adopt a novel approach showing the exact equality of the Bott index of two unitary matrices, given by \eqref{U_un} below with the Chern number. The invertible matrices  \eqref{U} and the unitary matrices  \eqref{U_un}, are shown to be related by a homotopy within the invertible matrices, implying the equality of their Bott indices. In subsubsection \ref{Bott-Chern-exact} I also discuss subtleties concerning homotopies and periodic boundary conditions on the lattice. 

According to sections 2F and 4 of \cite{Bellissard:1994}, the wordings ``Chern number'' and ``transverse conductance'' will be used in here as synonymous.

In the following the notation is that of section \ref{section_physical_setting}.

\subsection{The Hastings-Loring approach, and more} \label{Bott-Chern-approx}

Hastings and Loring considered in \cite{Hastings_Loring:2011} the pair of almost unitary matrices $ P\e^{\left(i\frac{2\pi X}{L}\right)}P $ and $ P\e^{\left(i\frac{2\pi Y}{L}\right)}P $ as acting on the subspace $ \textrm{Ran}(P) $.
I will consider instead the pair of almost unitary matrices over $ l^2(\Lambda) $, already introduced by Loring in section 9 of \cite{Loring_2015} given by \eqref{U} below. 

\begin{Lemma}
Given a Hamiltonian $ H $ as in \ref{Ham_single}, with $ P $ the Fermi projection, $ P = \chi(H\le \mu)$, and defining $ \theta_x:=\frac{2\pi X}{L} $, $ \theta_y:=\frac{2\pi Y}{L} $ and $ P^\bot := \mathds{1}-P $, the matrices  
\begin{align}
  P^\bot + P\e^{ i \theta_x }P \hspace{2mm}, P^\bot + P\e^{ i \theta_y }P \label{U}
\end{align}
are almost unitary, namely it holds:
\begin{align}
 & \| (P^\bot + Pe^{i\theta_x}P)^* (P^\bot + Pe^{i\theta_x}P) - \mathds{1} \| \ll 1 \\
& \| (P^\bot + Pe^{i\theta_x}P) (P^\bot + Pe^{i\theta_x}P)^* - \mathds{1} \| \ll 1
\end{align}
The same is true replacing $ \theta_x $ with $ \theta_y $. Moreover the pair \eqref{U} almost commute:
\begin{align}
 \| [ \PP + Pe^{i\theta_x}P,  \PP + Pe^{i\theta_y}P ] \|  \ll 1
\end{align}
\end{Lemma}

\begin{proof}
First of all, we observe that it is easy to show that if $ A $ is such that $ \|A^*A-\I\| = \mathcal{O}(\lambda) $, with $\lambda \ll 1 $, then this implies $ \|AA^*-\I\| = \mathcal{O}(\lambda) $.
\begin{align}
 & \| (P^\bot + Pe^{i\theta_x}P)^* (P^\bot + Pe^{i\theta_x}P) - \mathds{1} \| = \| P^\bot + Pe^{-i\theta_x}Pe^{i\theta_x}P - \mathds{1} \| \\
 & = \|   Pe^{-i\theta_x}Pe^{i\theta_x}P -P\| = \| P(e^{-i\theta_x}(P-\I)e^{i\theta_x})P \| \\
 &= \| Pe^{-i\theta_x}P^\bot e^{i\theta_x}P \| = \| [P,e^{-i\theta_x}]P^\bot [e^{i\theta_x},P] \| \\
 &\le \| [P,e^{-i\theta_x}] \|^2 \le \mathcal{O} \left( \frac{R}{L}\frac{\|H \|}{\Delta E} \right)^2 \ll 1 \label{normproduct}
\end{align}
This implies $ \| (P^\bot + Pe^{i\theta_x}P) (P^\bot + Pe^{i\theta_x}P)^* - \mathds{1} \| \ll 1 $.
Moreover:
\begin{align}
 &\| [ \PP + Pe^{i\theta_x}P,  \PP + Pe^{i\theta_y}P ] \| = \| [ Pe^{i\theta_x}P, Pe^{i\theta_y}P ] \| \\
 &= \| P( e^{i\theta_x}Pe^{i\theta_y} - e^{i\theta_y}Pe^{i\theta_x})P ] \| = \| P( e^{i\theta_y}\PP e^{i\theta_x} - e^{i\theta_x}\PP e^{i\theta_y})P ] \| \\
& = \|[P, e^{i\theta_y}]\PP [e^{i\theta_x},P] - [P, e^{i\theta_x}]\PP [e^{i\theta_y},P]\| \le 2 \| [P,e^{i\theta_x}] \| \|[P,e^{i\theta_y}] \| \label{commxy}\\
& \le \mathcal{O} \left( \frac{R}{L}\frac{\|H \|}{\Delta E} \right)^2 \ll 1
\end{align}
\end{proof}

We now want to evaluate the Bott index of the pair \eqref{U}. With the aid of equation \eqref{approxlog} below, already stated in section 5.3 of \cite{Hastings_Loring:2011}, we get rid of the $ \log $ in equation \eqref{Bott_trace} introducing a $ \mathcal{O}(\lam^{2}) $ correction, with $ \lam:= \frac{R \|H\|}{L \, \Delta E} $.
\begin{Lemma}
 Given $ U $ and $ V $ two invertible matrices over $ \mathds{C}^N $, $ N \gg 1 $, and given a parameter $ g $ with $ g^2 \propto \frac{1}{N} $, with $ U $ and $ V $ satisfying:
 \begin{align}
 & \| U^* U - \mathds{1} \| = \mathcal{O}(g^{2}), \hspace{5mm} \| U U^* - \mathds{1} \| = \mathcal{O}(g^{2})  \\
 & \| V^* V - \mathds{1} \| = \mathcal{O}(g^{2}), \hspace{5mm} \| V V^* - \mathds{1} \|  = \mathcal{O}(g^{2}) \\
& \| [U,V] \| = \mathcal{O}(g^{2})
\end{align}
namely they are almost unitary and they almost commute. 
This implies that:
\begin{eqnarray}
 &\|\log(UVU^{-1} V^{-1}) - (UVU^{-1} V^{-1}-\I) \|\le \mathcal{O}(g^{4}) \label{approxlog}  \\
 & | \mathrm{Bott}(U,V)-\frac{1}{2\pi}\mathrm{Im}\mathrm{Tr}(UVU^{-1} V^{-1}) | \le \mathcal{O}(g^{2}) \label{Bott_trace}
\end{eqnarray}
\end{Lemma}

\begin{proof}
Let us consider the $ \log $ series, given  $ A $ with 
$ \|A-\id \|<1 $, that implies $ A $ invertible, it holds:  
\begin{equation}
\log A = \sum_{n=1}^ \infty (-1)^{n+1} \frac{(A-\id)^n}{n} 
\end{equation}
Then: 
\begin{align}
 \| \log A - (A-\id) \| &= \| \sum_{n=2}^ \infty (-1)^{n+1} \frac{(A-\id)^n}{n}\|  \le \|A-\id \|^2   \sum_{n=2}^ \infty  \frac{\|A-\id \|^{n-2}}{n}  \label{four} 
\end{align}
Equation \eqref{approxlog} follows from \eqref{four} with $ A = UVU^{-1} V^{-1} $.  The modulus of the trace of a matrix is less equal than the trace norm that is less equal than the 
norm of the matrix itself times the dimension of the space the matrix is acting upon, then:
\begin{equation}
 | \textrm{Tr}[ \log(UVU^{-1} V^{-1}) - (UVU^{-1} V^{-1} - \I)] | \le \mathcal{O}(g^{4}) \mathcal{O}(N) = \mathcal{O}(g^{2})
\end{equation}
Noticing that 
\begin{align}
 \mathrm{Bott}(U,V) = \frac{1}{2\pi}\mathrm{Im} \mathrm{Tr} \log (UVU^{-1} V^{-1}) 
\end{align}
equation \eqref{Bott_trace} follows.
\end{proof}

To show the equality of the Bott index of the pair \eqref{U} with the Chern number of $ P $ we make use of an expression of the latter that is suitable for this proof. The name Chern number arises within the theory of Chern class, see \cite{DeWitt_Morette_1,Dubrovin_1}, as an invariant of manifolds. For infinite systems in dimension two when the Hilbert space of the system is  $ l^2(\mathds{Z}^2) $ the authors of \cite{Bellissard:1994} have shown that given a Hamiltonian with a local Fermi projection $ P $, $ P\frac{X+iY}{|X+iY|}P + \PP $, is a Fredholm operator, with index equal, after average over the disorder distribution, to the transverse conductance. The same authors in section 2F and section 4 of their work \cite{Bellissard:1994} develop the linear response theory that provides the form of the transverse conductance, this coincides with the Chern character (number) of the projection $ P $ defined as follows: 
\begin{equation} \label{def_Ch}
 \mathrm{Ch}(P)= -2 \pi i \Tr_{u. a.} P \left[\partial_x P,\partial_y P \right]
\end{equation}
$ \partial_x $ and $ \partial_y $ denote non commutative derivatives, namely $ \partial_x A := [-iX, A] $, $ \partial_y A := [-iY, A] $, with $ X $ and $ Y $ the position operators. 
The trace $ \mathrm{Tr_{u. a.}} $ stays for the trace per unit area, that is:  $ \mathrm{Tr_{u. a.}} := \lim_{A \rightarrow \infty} \frac{\mathrm{Tr_A}}{A}$.  
From \eqref{def_Ch} it follows that 
\begin{equation} \label{Chernnumber}
\mathrm{Ch}(P) = -4\pi \mathrm{Im} \mathrm{Tr_{u. a.}}P \left[X,P\right] [Y,P]
\end{equation}
In fact:
\begin{align}
 & 2\pi i\Tr_{u. a.} P \left[\partial_x P,\partial_y P \right] =2 \pi  i\Tr_{u. a.} P \left[ [-iX, P] , [-iY, P] \right] = \\
 & = 2 \pi i\Tr_{u. a.} P \left( [-iX, P]  [-iY, P] - [-iY, P][-iX, P] \right) \label{proj_with_P} =
 -4 \pi \mathrm{Im} \mathrm{Tr_{u. a.}}P \left[-iX,P\right]  [-iY,P]
\end{align}
I remark that in here I follow a sign convention for the Chern number different from \cite{Bellissard:1994}, but in agreement with definitions 6.3 and 6.6 of \cite{Avron_Seiler_Simon:1994}, and in agreement also with appendix C of \cite{Kitaev:2009}.
The definition of Chern number also generalizes to  higher dimensions as can be seen in chapter 6 of \cite{Prodan_Schulz_Baldes}.
Prodan has discussed in theorem 5.11 and corollary 5.12 of \cite{Prodan_review:2011} the stability properties of the Chern number with respect to deformations of the Hamiltonian that gives rise to the Fermi projection $ P $.

As far as regards the physical system on the torus that we are considering, as described in section \ref{section_physical_setting}, the definition of Chern number is as in equation \eqref{Chernnumber} with $ \mathrm{Tr_{u. a.}} (\cdot)=  \frac{1}{L^2} \mathrm{Tr_{l^2(\Lambda)}} (\cdot) $.

\begin{Theorem} \label{HL_approach}

Given a Hamiltonian $ H $ as in definition \ref{Ham_single} and its Fermi projection $ P $, the approximated expression of the Bott index, as in eq. \eqref{Bott_trace}, of the almost unitary matrices $  P^\bot + P\e^{\left(i\frac{2\pi X}{L}\right)}P $ and $  P^\bot + P\e^{\left(i\frac{2\pi Y}{L}\right)}P $  equals the 
Chern number of the projection $ P $, as given in equation \eqref{Chernnumber}, within a correction of order $ O\left( \frac{R \|H\|}{L \, \Delta E} \right)$.
\begin{equation} \label{approx_eq}
\mathrm{Bott}\left(P^\bot + P\e^{\left(i\frac{2\pi X}{L}\right)}P,P^\bot + P\e^{\left(i\frac{2\pi Y}{L} \right)}P \right) = \mathrm{Ch}(P) + \mathcal{O}(\lambda) 
\end{equation}
with $ \lambda := \frac{R \|H\|}{L \, \Delta E} $.
\end{Theorem}

\begin{proof} 
With an eye towards operators in infinite dimensional Hilbert spaces, it is a good idea to have a control on the order of magnitude of a trace, therefore starting from equation \eqref{square} all the terms appearing under a trace are of order $ L^{-2} $, being in our model the Hilbert space of dimension of order $ L^2 $, it means that we are handling traces of $\mathcal{O}(1) $.
\begin{align}
&\mathrm{Bott}\left(P^\bot + P\e^{\left(i\frac{2\pi X}{L}\right)}P,P^\bot + P\e^{\left(i\frac{2\pi Y}{L} \right)}P \right)= \nonumber \\
&=\frac{1}{2\pi}\mathrm{Im}\mathrm{Tr} \left( \I + [U,V]U^{-1}V^{-1} \right) + \mathcal{O}(\lam^{2}) =\frac{1}{2\pi}\mathrm{Im}\mathrm{Tr}\left( \I+ [U,V]U^*V^* \right) + \mathcal{O}(\lam^{2}) \\
&=\frac{1}{2\pi}\mathrm{Im}\mathrm{Tr}\left( [P\e^{i\theta_x} P,P\e^{i\theta_y} P]P\e^{-i\theta_x} P\e^{-i\theta_y} P \right) + \mathcal{O}(\lam^{2}) \label{square} \\
&=\frac{1}{2\pi}\mathrm{Im}\mathrm{Tr} \left[ \left( P\e^{i\theta_y} \PP \e^{i\theta_x} P - \e^{i\theta_x} \PP \e^{i\theta_y} P\right) \e^{-i\theta_x} P\e^{-i\theta_y} P \right]  + \mathcal{O}(\lam^{2})  \\
&=\frac{1}{2\pi}\mathrm{Im}\mathrm{Tr} \left( P \e^{-i\theta_y}P\e^{i\theta_y} \PP \e^{i\theta_x} P \e^{-i\theta_x} P - P \e^{-i\theta_y}P\e^{i\theta_x} \PP \e^{i\theta_y} P \e^{-i\theta_x} P \right) + \mathcal{O}(\lam^{2})  \label{61}
\end{align}
The terms inside the trace in equation \eqref{61} have quite a similar structure:
\begin{align}
 &P\e^{-i\theta_y} P\e^{i\theta_y} \PP \e^{i\theta_x}P\e^{-i\theta_x}P = P[-i\theta_y, P]\PP[i\theta_x, P]P + \mathcal{O}(\lambda)^3 \label{62}
\end{align}
Inserting a couple of identities $ \I=\e^{i\theta_y}\e^{-i\theta_y}=\e^{-i\theta_x}\e^{i\theta_x}$ in the second term of \eqref{61} we get:
\begin{align}
& P \e^{-i\theta_y}P\e^{i\theta_x} \PP \e^{i\theta_y} P \e^{-i\theta_x} P= 
P \e^{-i\theta_y}  P \e^{i\theta_y}\e^{-i\theta_y}  \e^{i\theta_x} \PP \e^{i\theta_y} \e^{-i\theta_x}\e^{i\theta_x} P \e^{-i\theta_x} P  \\
&= P\left( P + [-i\theta_y,P] \right) ( \PP + [i(\theta_x - \theta_y), \PP] ) \left( P + [i\theta_x,P] \right)P + \mathcal{O}(\lambda)^3 \label{64}  \\
&= -P [i\theta_x , P] \PP [i\theta_x,P] P + P [i\theta_y , P] \PP [i\theta_x,P] P - P [-i\theta_y , P] \PP [i\theta_x,P] P+ \nonumber \\
& + P [-i\theta_y , P] \PP [i\theta_y,P]P + P [-i\theta_y , P] \PP [i\theta_x,P]P + \mathcal{O}(\lambda)^3 \label{63}
\end{align}
In equation \eqref{63} there are two Hermitean terms, therefore they  have imaginary part of the trace vanishing, and two other terms cancel.
Going back to \eqref{61}, we get:
\begin{align}
&\mathrm{Bott}\left(P^\bot + P\e^{\left(i\frac{2\pi X}{L}\right)}P,P^\bot + P\e^{\left(i\frac{2\pi Y}{L} \right)}P \right)= \frac{1}{\pi}\mathrm{Im}\mathrm{Tr} \left(P[\theta_y,P]\PP [\theta_x,P] \right) + \mathcal{O}(\lam) \\
&  = -\frac{1}{\pi}\mathrm{Im}\mathrm{Tr} \left( P[ \theta_x,P][\theta_y,P] \right) + \mathcal{O}(\lam) = -\frac{4\pi}{L^2}\mathrm{Im}\mathrm{Tr} \left(P[ X,P][ Y ,P] \right) + \mathcal{O}(\lam) \label{Bott_final}
\end{align}
We see  that equation \eqref{Bott_final} coincides with equation \eqref{Chernnumber} up to $\mathcal{O}(\lam)$.
Some remarks from the equations above: equation \eqref{62} and \eqref{64} follows from the identity, with $ A $ Hermitean:
\begin{align}
 e^{iA}Be^{-iA}-B=[iA,B]+\frac{1}{2}[iA,[iA,B]]+ \ldots + \frac{1}{n!}[iA \ldots [iA,B] \ldots ] + \ldots
\end{align}
$ \|[\theta_x, P] \| \le \mathcal{O}(\lam) $ is equation 5.5 of \cite{Hastings_Loring:2010}, this has been used in \eqref{62} and \eqref{64}. This also follows from the application of the Holmgren bound, in a similar fashion to what done in Lemma \ref{lemma_3}, to $ \| [X,H] \| $, with $ |\langle m |[X,H]|n\rangle|=\textrm{dist}(m_x,n_x)|\langle m |H|n\rangle| $. With $ m_x $ and $ n_x $ in the set $ [0,L-1] \cap \mathds{Z} $ the distance $ \textrm{dist} $ reflects the periodic boundary conditions, namely
\begin{equation}
 \textrm{dist}(m_x,n_x) = \min \{ |n_x - m_x|,L-|n_x - m_x| \} 
\end{equation}
Application of the same ideas leads to $ \frac{1}{L^2} \| [X,[X,H]] \| =  \mathcal{O}(\lam^2) $.    

The first equation in \eqref{Bott_final} follows from:
\begin{align}
 &\mathrm{Im}\mathrm{Tr} \left(P[\theta_y,P]\PP [\theta_x,P] \right)= \mathrm{Im}\mathrm{Tr} \left([\theta_x,P] [\theta_y,P] \PP \right) = -\mathrm{Im}\mathrm{Tr} \left(P[\theta_x,P] [\theta_y,P] \right) \label{71}
\end{align}
In \eqref{71} it has been used: $ [A,P]=P[A,P]\PP+\PP[A,P]P$ with $ A $ bounded and $ P $ a projection, this implies that $ P[A,P]P=0 $. It has also been used: given $ B $ and $ C $ skew adjoint matrices then $ \mathrm{Im}\mathrm{Tr} (BC) = -\mathrm{Im}\mathrm{Tr} (BC)^* =  -\mathrm{Im}\mathrm{Tr} (C^*B^*) = -\mathrm{Im}\mathrm{Tr} (CB) =-\mathrm{Im}\mathrm{Tr} (BC) $, implying $ \mathrm{Im}\mathrm{Tr} (BC) =0 $.

We stress that $ [X,P] $ is well defined with periodic boundary conditions but neither $ XP $ nor $ PX $ is well defined, if singularly taken. This implies that \eqref{Bott_final} is well defined on a finite torus and coincides up to corrections of order $ \mathcal{O}(\lam) $ with the transverse conductance.
\end{proof}

Hastings and Loring have developed the theory of the linear response on the torus in section 5.3 of \cite{Hastings_Loring:2011} making use of the current operator, with $ \theta_x=2\pi\frac{X}{L}$:
\begin{equation}
J_x= \frac{1}{2}\left(e^{i\theta_x}He^{-i\theta_x}-e^{-i\theta_x}He^{i\theta_x}\right) = \frac{2\pi}{L}[iX,H] + \mathcal{O}(L^{-3})
\end{equation}
They show that the Bott index equals the transverse conductance on the torus up to correction of order $ \mathcal{O}(L^{-1}) $. 

It is worth mentioning that the Chern number of a finite dimensional projection $ P $, $ \textrm{rank}(P) < \infty $, defined on an infinite dimensional Hilbert space or on a finite dimensional Hilbert space with open boundary conditions is vanishing, in fact:
\begin{align}
 &\mathrm{Im}\mathrm{Tr} \left(P[X,P][Y ,P] \right) = \mathrm{Im}\mathrm{Tr} \left(P[X,P]\PP[Y ,P] \right) \label{first} \\
 &= \mathrm{Im}\mathrm{Tr} \left( PX\PP YP \right) = \mathrm{Im}\mathrm{Tr} \left( PX YP -  PX PYP  \right) \label{second} \\
 &=-\mathrm{Im}\mathrm{Tr} \left( PX PYP  \right) = -\mathrm{Im}\mathrm{Tr} \left( PY PXP  \right) =0 \label{0_Chern}
\end{align}
In the first equality of \eqref{0_Chern} I have used the cyclic property of the trace with respect the two blocks $ PXP $, $PYP$; in the second equality the fact that if the trace of a (trace class) operator coincides with the trace of its adjoint then it is real. This has also been used in \eqref{second}. 

On the torus, namely with periodic boundary conditions, we cannot ``open'' the commutator and take the trace of each operator, like it has been done in \eqref{second}, in fact in that case $  PX YP $ and $PX PYP$ are not well defined with respect to periodic boundary conditions.

A proof of the Bott index - Chern number equivalence based on a ``momentum space'' approach has 
been discussed in \cite{Ge_Rigol:2017}. Also a version of the Bott index in momentum space (despite no definition was given) can be spotted reading among the lines of \cite{Hatsugai_2005}.

\subsection{An exact approach. Homotopies on the torus: the right and the wrong way} \label{Bott-Chern-exact}

As stated at the beginning of section \ref{section_equivalence}, I will provide here a novel proof of the Bott index - Chern number correspondence, showing the equality among the Bott index of the pair of unitary matrices given below in equation \eqref{U_un} and the Chern number $ \mathrm{Ch}(P) $ taking advantage of an approach developed in \cite{Toniolo_Bott_2021} for bounded operators. The operators \eqref{U_un} have been already suggested in the context of the integer quantum Hall effect by Kitaev in the appendix C of \cite{Kitaev_2006}, and more recently, with a suitable modification, in the context of infinite dimensional Hilbert spaces by the authors of \cite{Shapiro_2020, Bols_2021}.
Being the Bott index an integer this shows that the correction $ \mathcal{O}(\lambda) $ in equation \eqref{approx_eq} is actually vanishing. As an application of the homotopy invariance of the Bott index I will also show in lemma \ref{homotopy} below that the two almost unitary matrices \eqref{U} and the two unitary matrices \eqref{U_un} are connected by a homotopy within the invertible matrices, providing another proof of the equality of their Bott indices.

\begin{Theorem} \label{theo2}
Given a Hamiltonian $ H $ as in definition \ref{Ham_single} and its Fermi projection $ P $, the unitary matrices
\begin{align}
& e^{2 \pi i P \frac{X}{L} P } \hspace{2mm} , \hspace{2mm} e^{2 \pi i P \frac{Y}{L} P } \label{U_un}
\end{align}
 have a well defined Bott index that satisfies: 
\begin{equation} \label{Bott-Chern}
 \textrm{Bott}\left( e^{2 \pi i P \frac{X}{L} P },e^{2 \pi i P \frac{Y}{L} P } \right) = 2\pi i \mathrm{Tr} \left[ P \frac{X}{L}P,P \frac{Y}{L}P \right] = -\frac{4\pi}{L^2}\mathrm{Im}\mathrm{Tr} \left(P[X,P][Y ,P] \right)= \mathrm{Ch}(P)
\end{equation}
\end{Theorem}

\begin{proof}
First of all we start noticing that $ e^{2 \pi i P \frac{X}{L} P }=Pe^{2 \pi i P \frac{X}{L} P }P+\PP$, implying that $ [P,e^{2 \pi i P \frac{X}{L} P }]=0$.
Let us show that the unitary matrices in \eqref{U_un} almost commute implying that their Bott index is well defined. 
This is implied by $ \| P^\bot + P e^{i 2\pi \frac{ X}{L}}P - e^{2 \pi i P \frac{X}{L} P} \| \le \mathcal{O}(\lam^2) $ that can be shown as follows:
\begin{align}
P^\bot + P e^{i 2\pi \frac{ X}{L}}P - e^{2 \pi i P \frac{X}{L} P} &= 
 P\left( e^{2 \pi i \frac{X}{L}} - e^{2 \pi i P \frac{X}{L} P }  \right)P \label{_75}\\ 
 &= P\left(\sum_{n=2}^\infty \frac{(2\pi i)^n}{n!}\left[ \left(\frac{X}{L}\right)^n - \left(P\frac{X}{L}\right)^n  \right] \right)P \label{P_diff}
\end{align}
Using the following equality of bounded operators $ A $ and $ B $
\begin{equation} 
 A^n-B^n= \sum_{j=0}^{n-1} B^j(A-B)A^{n-j-1} 
\end{equation}
We obtain with $ n \ge 2 $
\begin{align}
 P\left[ \left(\frac{X}{L}\right)^n - \left(P\frac{X}{L}\right)^n \right]P &= P\left[ \sum_{j=0}^{n-1} \left(P\frac{X}{L}\right)^j\left(\frac{X}{L}-P\frac{X}{L}\right)\left(P\frac{X}{L}\right)^{n-j-1} \right]P\\
 &=P\left[\sum_{j=0}^{n-1} \left(P\frac{X}{L}\right)^j \PP\frac{X}{L} \left(P\frac{X}{L}\right)^{n-j-1} \right]P\\
 &=P\left[\sum_{j=1}^{n-1} \left(P\frac{X}{L}\right)^j \PP\frac{X}{L} \left(P\frac{X}{L}\right)^{n-j-1} \right]P\\
 &=P\left[\sum_{j=1}^{n-1} \left(P\frac{X}{L}\right)^{(j-1)} P\frac{X}{L} \PP\frac{X}{L} P\left(P\frac{X}{L}\right)^{n-j-1} \right]P \\
 &=P\left[\sum_{j=1}^{n-1} \left(P\frac{X}{L}\right)^{(j-1)} \left[P,\frac{X}{L}\right] \PP \left[\frac{X}{L},P \right] \left(P\frac{X}{L}\right)^{n-j-1} \right]P \label{sum_to_n}
\end{align}
The norm of the matrix in equation \eqref{sum_to_n} is bounded by $ (n-2)\mathcal{O}(\lam^2) $. Using the series expansion of the exponential it  is possible to show that the norm of \eqref{P_diff} is bounded by $ \mathcal{O}(\lam^2) $.
In a more concise way we can obtain an upper bound of $\mathcal{O}(\lam)$  for \eqref{_75}, as follows:
\begin{align} \label{75}
 \| P^\bot + P e^{ i 2\pi \frac{X}{L} }P - e^{2 \pi i P \frac{X}{L} P} \| &= \| P \left( e^{ i\frac{2\pi X}{L}} - e^{2 \pi i P \frac{X}{L} P} \right)P \| \\
 &=\| 2\pi i P \int_0^1 e^{ i 2\pi \frac{ X}{L} t}\left(\frac{X}{L}-P\frac{X}{L}P\right) e^{2 \pi i P \frac{X}{L} P(1-t)}dt P \| \label{76} \\
 &=\| 2\pi i P \int_0^1 e^{ i 2\pi \frac{ X}{L} t}\left(\frac{X}{L}-P\frac{X}{L}P\right) P e^{2 \pi i P \frac{X}{L} P(1-t)}dt  \| \\
 & \le 2\pi \| \frac{X}{L}P-P\frac{X}{L}P\| = 2\pi \| P^\bot \frac{X}{L}P\| = 2\pi \| P^\bot \left[ \frac{X}{L},P \right]\| \\
 & \le \mathcal{O}(\lam)
\end{align}
In equation \eqref{76} a DuHamel formula has been used. 
Then:
\begin{align}
 &\| \left[ e^{2 \pi i P \frac{X}{L} P},e^{2 \pi i P \frac{Y}{L} P} \right] \| \\
 &= \| P \left[ e^{2 \pi i P \frac{X}{L} P} - P e^{ i 2\pi \frac{X}{L} }P +  P e^{ i 2\pi \frac{X}{L} }P ,e^{2 \pi i P \frac{Y}{L} P} - P e^{ i 2\pi \frac{Y}{L} }P + P e^{ i 2\pi \frac{Y}{L} }P\right] P\| \\
 & \le \|  \left[ P e^{ i 2\pi \frac{X}{L} }P , P e^{ i 2\pi \frac{Y}{L} }P \right] \| + \mathcal{O}(\lam)  \le \mathcal{O}(\lam) 
\end{align}
To prove equation \eqref{Bott-Chern} we consider the maps of unitary matrices $ U(t):[0,1] \rightarrow e^{2 \pi i t P \frac{X}{L} P } $ and $ V(t):[0,1] \rightarrow e^{2 \pi i t P \frac{Y}{L} P } $. With $ t \in (0,1) $, $U(t) $ and $ V(t) $ do not satisfy periodic boundary conditions therefore they are not admissible homotopies, according to lemma \ref{hom_invariance}, meaning that along the paths $U(t) $ and $ V(t) $, $ \mathrm{Bott}\left(U(t),V(t)\right)$ is allowed to change. This can be seen for example looking at equation \eqref{hom_inv}, we see that $ U^{-1}(t)\partial_t U(t) $ must be a well defined matrix over the given Hilbert space, in our case $ l^2(\Lambda) $ periodic boundary conditions. On the contrary we see that  
\begin{equation}
  U^{-1}(t)\partial_t U(t) =  2\pi iP \frac{X}{L}P.
\end{equation}

We now map the problem of determining the LHS of equation \eqref{Bott-Chern} into the solution of a first order different equation.

To simplify the notation we denote $ \phi_x:=2\pi P \frac{X}{L}P$ and $ \phi_y:=2\pi P \frac{Y}{L}P$.
Let us define $ g(t):= \T \log \left(e^{it\tx}e^{it\ty}e^{-it\tx}e^{-it\ty} \right)$;  the argument of the $ \log $ satisfies periodic boundary conditions, in fact with $ X \rightarrow X + nL\I $ and $ Y \rightarrow Y + mL\I $, with $ n $ and $ m \, \in \mathds{Z} $, we have that
\begin{align}
 e^{2\pi it P \left(\frac{X}{L}+n \I \right)P} &= e^{2\pi it n P}e^{2\pi it P \frac{X}{L}P} \\
 e^{2\pi it P \left(\frac{Y}{L}+m \I \right)P} &= e^{2\pi it m P}e^{2\pi it P \frac{Y}{L}P} 
\end{align}
This implies that the trace defining $ g(t) $ is well posed. We also notice that $ \|[e^{it\tx},e^{it\ty}]\| < 2 $.
\begin{align}
 \frac{dg}{dt}&=\T [(i\tx e^{it\tx}e^{it\ty}e^{-it\tx}e^{-it\ty}+e^{it\tx}i\ty e^{it\ty}e^{-it\tx}e^{-it\ty}-e^{it\tx}e^{it\ty}i\tx e^{-it\tx}e^{-it\ty} \\ &\hspace{10mm} -e^{it\tx}e^{it\ty}e^{-it\tx}i\ty e^{-it\ty})  e^{it\ty}e^{it\tx}e^{-it\ty}e^{-it\tx} ] \\
 &=\T \left(i\tx +e^{it\tx}i\ty e^{-it\tx}-e^{it\tx}e^{it\ty}i\tx e^{-it\ty}e^{-it\tx}-e^{it\tx}e^{it\ty}e^{-it\tx}i\ty e^{it\tx}e^{-it\ty}e^{-it\tx}\right) \nonumber \\
 &=\T \left(i\tx + i\ty -e^{it\ty}i\tx e^{-it\ty}-e^{it\ty}e^{-it\tx}i\ty e^{it\tx}e^{-it\ty}\right) \\
 &=\T \left(e^{-it\ty}i\tx e^{it\ty}+i\ty-i\tx-e^{-it\tx}i\ty e^{it\tx}\right) \\
 &=\T \left(e^{-it\ty}i\tx e^{it\ty} -i\tx \right) + \T \left( i\ty-e^{-it\tx}i\ty e^{it\tx} \right) \\
 &=\T \left( \int_0^1 ds e^{-ist\ty}[-it\ty,i\tx]e^{ist\ty} \right) + \T \left( \int_0^1 ds e^{-ist\tx}[i\ty,-ti\tx]e^{ist\tx} \right) \\
 &=\T \left( [-it\ty,i\tx] \right) + \T \left( [i\ty,-it\tx] \right) = 2t \T \left( [i\tx,i\ty] \right) 
\end{align}
Observing that $ g(0)=0$
\begin{align}
 g(t)=\int_0^t g'(s)ds=\int_0^t 2s\T[i\tx,i\ty] ds = t^2\T[i\tx,i\ty] 
\end{align}
This implies that
\begin{equation}
 \B(e^{i\tx},e^{i\ty})=\frac{1}{2\pi i}g(1)=\frac{1}{2\pi i} \T[i\tx,i\ty]
\end{equation}
Then 
\begin{equation} \label{Bott_exact}
\textrm{Bott}\left( e^{2 \pi i P \frac{X}{L} P },e^{2 \pi i P \frac{Y}{L} P } \right) = 2\pi i \mathrm{Tr} \left[ P \frac{X}{L}P,P \frac{Y}{L}P \right]
\end{equation}
Moreover 
\begin{align}
 &2\pi i \mathrm{Tr} \left[ P \frac{X}{L}P,P \frac{Y}{L}P \right] = \frac{2\pi i}{L^2} \mathrm{Tr} \left( P XPYP -PYPXP \right) \\
 &=\frac{2\pi i}{L^2} \mathrm{Tr} \left( PY\PP XP -PX \PP Y P \right) = \frac{4\pi }{L^2} \mathrm{Im} \mathrm{Tr}  PX \PP Y P \\
 &= \frac{4\pi }{L^2} \mathrm{Im} \mathrm{Tr}  [P,X] \PP [Y, P] = -\frac{4\pi }{L^2} \mathrm{Im} \mathrm{Tr}  P[X,P]  [Y, P] \label{final}
\end{align}
\end{proof}

\begin{Lemma} \label{homotopy}
Given $ H $ as in definition \ref{Ham_single} and $ P $ its Fermi projection, it holds: 
\begin{equation} \label{equality}
 \textrm{Bott}\left(P e^{2 \pi i  \frac{X}{L}  }P + \PP , Pe^{2 \pi i  \frac{Y}{L} }P + \PP \right)  = \textrm{Bott}\left( e^{2 \pi i P \frac{X}{L} P },e^{2 \pi i P \frac{Y}{L} P } \right) 
\end{equation}
\end{Lemma}
\begin{proof}
The equation \eqref{equality} follows from the relation with the Chern number established in theorems \ref{HL_approach} and \ref{theo2}. To exemplify the construction of a homotopy, equation \eqref{equality} can also be proven considering the paths:
\begin{align}
 &\rho(s):=(1-s)e^{2 \pi i P \frac{X}{L} P }+s\left(P e^{2 \pi i \frac{X}{L}}P + \PP \right) \\
 &\eta(s):=(1-s)e^{2 \pi i P \frac{Y}{L} P}+s\left(P e^{2 \pi i \frac{Y}{L}}P + \PP \right)
\end{align}
with $ s \in [0,1] $. It is important to stress that $ \rho(s) $ and $ \eta(s) $ are well defined for all $ s \in [0,1] $ with respect to periodic boundary conditions, namely $ \rho(s) $ is invariant when $ X \rightarrow X + n L $, $ \forall n \, \in \mathds{Z} $. 
$ \rho(s) $ is almost unitary: $ \rho^*(s)\rho(s) = \I + \mathcal{O}(\lam)$, that implies $ \rho(s)\rho^*(s) = \I + \mathcal{O}(\lam)$.  $ \rho^*(s)\rho(s) = \I + \mathcal{O}(\lam)$ can be seen as follows:
\begin{align}
 \rho(s)=(1-s)e^{2 \pi i P \frac{X}{L} P }+s\left(P e^{2 \pi i \frac{X}{L}}P + \PP \right) = e^{2 \pi i P \frac{X}{L} P } + sP\left( e^{2 \pi i \frac{X}{L}} - e^{2 \pi i P \frac{X}{L} P }  \right)P \label{hom_Chern}
\end{align}
then, according to equation \eqref{75}, the term proportional to $ s $ in \eqref{hom_Chern} is upper bounded by $ \mathcal{O}(\lam) $.
The same holds for $ \eta(s) $.

It is  $ \forall s \in [0,1] $, $ \|[\rho(s), \eta(s)]\| < 2 $.
Let us verify this. Setting as before $ \theta_x=2\pi \frac{X}{L} $ and $ \theta_y=2\pi \frac{X}{L} $, we have:
\begin{align}
 \|[\rho(s), \eta(s)]\|= &\|[(1-s)e^{i P \theta_x P }+s\left(P e^{ i  \theta_x  }P + \PP \right) , (1-s)e^{ i P \theta_y P }+s\left(P e^{ i  \theta_y  }P + \PP \right)] \| \nonumber \\
 & \le (1-s)^2 \| [e^{ i P \theta_x P},e^{ i P \theta_y P}]\| +  (1-s)s\|[e^{ i P \theta_x P},P e^{ i \theta_y} P] \| \nonumber \\
 & + (1-s)s\|[e^{ i P \theta_y P},P e^{ i \theta_x}P] \| + s^2\|[P e^{ i \theta_x}P,P e^{ i \theta_y}P] \| \\
 & \le (1-s)^2  \|[P \theta_x P,P \theta_y P] \| | +  (1-s)s\|[e^{ i P \theta_x P},e^{ i \theta_y}] \| \nonumber \\ 
 & + (1-s)s\|[e^{i P \theta_y P}, e^{ i \theta_x}] \| + s^2 \mathcal{O}\left(\lam^2 \right) \\
 & \le (1-s)^2   \mathcal{O}\left(\lam^2\right)  +  (1-s)s \|[ P \theta_x P, \theta_y] \| \nonumber \\
 & + (1-s)s \|[ P \theta_y P,\theta_x] \| + s^2 \mathcal{O}\left(\lam\right)^2 \\
 & \le (1-s)^2   \mathcal{O} (\lam^2)  + 2 (1-s)s \mathcal{O} (\lam) + s^2 \mathcal{O}(\lam^2) \ll 1 
\end{align}
\end{proof}
As a final remark about the subtleties of homotopies let us consider 
\begin{equation}
 W(s) := Pe^{i2 \pi (1-s\PP) \frac{X}{L} (1-s\PP)}P + \PP 
\end{equation}
that is an invertible map with the same initial and final point of $ \rho(s) $. Nevertheless $ W(s)$ does not satisfy periodic boundary conditions with $ s \in (0,1) $ therefore it is not admissible as a homotopy for the torus geometry.

%----------------------------------------------------------------
%                                       DISCUSSION
%----------------------------------------------------------------

\section{Discussion and perspectives} \label{section_discussion}

The formulation of the theory that leads to the construction of the Bott 
index is compatible with weak-disorder, meaning that disorder is admitted in the description of the system by the model Hamiltonian as far as a spectral gap is present. The role of the disorder in systems 
with topological features is essential in fact, for the integer quantum Hall effect for example, it is the presence of strong-disorder that 
makes possible the presence of plateaus in the shape of the Hall conductance as the external 
magnetic field is varied. See the introduction of reference \cite{Bellissard:1994} for a discussion. The 
Chern number admits a formulation developed in \cite{Bellissard:1994}   
that shows its quantization even in the presence of a disorder strong enough to close the spectral gap, for a visual illustration of this 
see figure 1 of reference \cite{Prodan_Hughes_Bernevig:2010}. 
The presence of a mobility gap, meaning that at the chemical potential there are only localized states, is still required, otherwise the system would lose its 
insulating nature. A different rigorous approach to strong-disordered systems is that of the so called deterministic disorder, see for example \cite{Simon_1995} and section 7 of \cite{del_Rio_1996} with the definition of SULE, and the more recent works \cite{Elgart_2005,Bols_2021}.  
In the formulation of the Bott index a spectral gap has been assumed. It seems natural to ask if the definition of the Bott index can be modified to accommodate for strong-disorder. 

Periodic driven Hamiltonians may host a peculiar topological invariant named, after \cite{Rudner_Levin_2013}, $ W $ invariant, that does not have a counterpart in the static case. The authors of \cite{Graf_Tauber_2018,Sadel_Schulz_Baldes_2017,Shapiro_Tauber_2019}, among others, have extended the definition of the $ W $ invariant to the weak- and strong-disordered cases for infinite systems in two dimensions. A formulation of the $ W $ invariant for finite and disordered systems, in particular the case of periodic boundary conditions considered here, as far as I know, is  missing. 

Finally I would like to offer a connection with the spectral localizer, see \cite{Schulz_Baldes_2019} and references therein. Is it possible to replace the Fermi projection with the Hamiltonian, or a computationally straightforward function of the Hamiltonian, in the evaluation of the Bott index, for example via homotopy? This would make the index faster to compute, also affording larger samples. The reference \cite{Loring_2019} provides a direct comparison among the two indices.

\section{Acknowledgements} 
I acknowledge financial support by the UK’s Engineering and Physical Sciences Research Council (grant number EP/R012393/1 Masanes).

\section{Data availability statement}
Data sharing not applicable to this article as no datasets were generated or analysed during the current study.

\section{Conflict of interest statement} 
The author states that there is no conflict of interest.

\section*{Appendices}
\appendix

\section{Holmgren bound} \label{appendix_Holmgren_bound}

$ A:\mathcal{H}\rightarrow \mathcal{H} $, a bounded operator over a separable Hilbert space $ \mathcal{H} $. It holds: 
\begin{equation}
 \|A\| \le \sqrt{\sup_m \sum_n |A_{n,m}|} \sqrt{\sup_n \sum_m |A_{n,m}|}
\end{equation}
$ A_{n,m} := \langle \chi_n,A \chi_m \rangle $, with $ \{ \chi_n \} $ any ONB of $ \mathcal{H} $.

\begin{proof}
$ \|A\| := \sup_{\{\|\phi\|=1,\|\psi\|=1\}} |\langle \phi, A \psi \rangle | $ 
 \begin{align}
  |\langle \phi, A \psi \rangle | & = |\sum_{n,m} \phi_n^* A_{n,m} \psi_m | \le \sum_{n,m} |\phi_n| |A_{n,m}| |\psi_m| = \sum_{n,m} |\phi_n| \sqrt{|A_{n,m}|} \sqrt{|A_{n,m}|} |\psi_m| \\
  & \le \sum_{n,m} |\phi_n| \left( \sup_m \sqrt{|A_{n,m}|} \right) \left( \sup_n \sqrt{|A_{n,m}|} \right) |\psi_m| \\
  & = \sum_{n} |\phi_n| \left( \sup_m \sqrt{|A_{n,m}|} \right) \sum_m \left( \sup_n \sqrt{|A_{n,m}|} \right) |\psi_m| 
 \end{align}
Applying  Cauchy-Schwarz to the sums on $ n $ and $ m $ considered each as a scalar product, we get:
\begin{align}
 |\langle \phi, A \psi \rangle | & \le \sup_m \sqrt{\sum_a |\phi_a|^2} \sqrt{\sum_b |A_{b,m}|} \sup_n \sqrt{\sum_c |A_{n,c}|} \sqrt{\sum_d |\psi_d|^2} 
 \end{align}
It follows that:
\begin{equation}
 \|A\| \le  \sup_m \sqrt{\sum_b |A_{b,m}|} \sup_n \sqrt{\sum_c |A_{n,c}|}   
\end{equation}

\end{proof}

\bibliography{biblio}

\end{document}